\documentclass[aps,prb,showpacs,twocolumn,citeautoscript,superscriptaddress]{revtex4}

\usepackage{graphicx}
\usepackage{bm}
\usepackage{amssymb}
\usepackage{amsmath}
\usepackage{dcolumn}
\usepackage{float}

\begin{document}

\title{Effect of Ni-doping on magnetism and superconductivity in Eu$_{0.5}$K$_{0.5}$Fe$_{2}$As$_{2}$}

\author{Anupam}
\email[]{anupam@iitk.ac.in}
\affiliation{Department of Physics, Indian Institute of Technology Kanpur, Kanpur-208016, India}

\author{V. K. Anand}
\affiliation{DCMP$\&$MS, Tata Institute of Fundamental Research, Mumbai-400005, India}

\author{P. L. Paulose}
\affiliation{DCMP$\&$MS, Tata Institute of Fundamental Research, Mumbai-400005, India}

\author{S. Ramakrishnan}
\affiliation{DCMP$\&$MS, Tata Institute of Fundamental Research, Mumbai-400005, India}

\author{C. Geibel}
\affiliation{Max Planck Institute for Chemical Physics of Solids, 01187 Dresden, Germany}

\author{Z. Hossain}
\affiliation{Department of Physics, Indian Institute of Technology Kanpur, Kanpur-208016, India}

\date{\today}

\begin{abstract}

The effect of Ni-doping on the magnetism and superconductivity in Eu$_{0.5}$K$_{0.5}$Fe$_{2}$As$_{2}$ has been studied through a systematic investigation of magnetic and superconducting properties of Eu$_{0.5}$K$_{0.5}$(Fe$_{1-x}$Ni$_{x}$)$_{2}$As$_{2}$ ($x$ = 0, 0.03, 0.05, 0.08 and 0.12) compounds by means of dc and ac magnetic susceptibilities, electrical resistivity and specific heat measurements. Eu$_{0.5}$K$_{0.5}$Fe$_{2}$As$_{2}$ is known to exhibit superconductivity with superconducting transition temperature $T_{c}$ as high as 33 K. The Ni-doping leads to a rapid decrease in $T_{c}$; $T_{c}$ is reduced to 23 K with 3\% Ni-doping, and 8\% Ni-doping suppresses the superconductivity to below 1.8 K. In 3\% Ni-doped sample Eu$_{0.5}$K$_{0.5}$(Fe$_{0.97}$Ni$_{0.03}$)$_{2}$As$_{2}$ superconductivity coexists with short range ordering of Eu$^{2+}$ magnetic moments at $T_{m}$ $\approx$ 6 K. The suppression of superconductivity with Ni-doping is accompanied with the emergence of a long range antiferromagnetic ordering with $T_{N}$ = 8.5 K and 7 K for Eu$_{0.5}$K$_{0.5}$(Fe$_{0.92}$Ni$_{0.08}$)$_{2}$As$_{2}$ and Eu$_{0.5}$K$_{0.5}$(Fe$_{0.88}$Ni$_{0.12}$)$_{2}$As$_{2}$, respectively. The temperature and field dependent magnetic measurements for $x$ = 0.08 and 0.12 samples reflect the possibility of a helical magnetic ordering of Eu$^{2+}$ moments. We suspect that the helimagnetism of Eu spins could be responsible for the destruction of superconductivity as has been observed in Co-doped EuFe$_{2}$As$_{2}$. The most striking feature seen in the resistivity data for $x$ = 0.08 is the reappearance of the anomaly presumably due to spin density wave transition at around 60 K. This could be attributed to the compensation of holes (K-doping at Eu-site) by the electrons (Ni-doping at Fe site). The anomaly associated with spin density wave further shifts to 200 K for $x$ = 0.12 for which the electron doping has almost compensated the holes in the system.

\end{abstract}


\maketitle

\section{INTRODUCTION}

The discovery of superconductivity  (SC) in AFe$_{2}$As$_{2}$ (A = Sr, Ba, Ca) upon suppressing the spin-density wave (SDW) transition by doping Fe-site with other transition metal elements is quite different from the high-$T_{c}$ cuprate superconductors where superconductivity was always suppressed by even small levels of substitution at the electronically active Cu site.\cite{1,2,3,4} Superconductivity has been observed in BaFe$_{2}$As$_{2}$ and SrFe$_{2}$As$_{2}$ by the substitution of 3d and 4d transition elements e.g. Ni, Co, Rh, Ru, Ir etc. at Fe site along with the coexistence of spin density wave transition and superconductivity in the under-doped samples.\cite{5,6,7,8,9,10,11,12,13,14} Among the 122-pnictide, EuFe$_{2}$As$_{2}$ is an interesting member, the magnetic Eu$^{2+}$ moments order antiferromagnetically below 19 K along with the spin density wave transition at 190 K.\cite{15} Due to the presence of Eu magnetic moments, partial substitution at Eu, Fe or As-sites leads to many interesting properties like the coexistence of magnetism and superconductivity as observed in Eu$_{1-x}$K$_{x}$Fe$_{2}$As$_{2}$\cite{16} and re-entrant superconductivity on application of hydrostatic and chemical pressure.\cite{17,18,19} In contrast to Co-doped BaFe$_{2}$As$_{2}$ and SrFe$_{2}$As$_{2}$, Co-doping in EuFe$_{2}$As$_{2}$ leads to re-entrant superconductivity below 21 K\cite{20} and Ni-doping does not show any superconducting transition down to 2 K.\cite{21} While the Eu moments are reported to order ferromagnetically below 20 K in Ni-doped EuFe$_{2}$As$_{2}$, a helical magnetic structure of Eu$^{2+}$ moments has been proposed for Eu(Fe$_{0.89}$Co$_{0.11}$)$_{2}$As$_{2}$.\cite{20} This helimagnetism in Co-doped EuFe$_{2}$As$_{2}$ destroys the superconductivity while the field induced ferromagnetism favours the superconductivity in the same system.\cite{20} M\"{o}ssbauer studies on these systems have revealed that the magnetic ground state of Fe sublattice strongly affects the direction of Eu moments.\cite{22} When the Fe ions are in SDW state, the direction of Eu magnetic moment is in the ab-plane, on the other hand in the absence of SDW ordering of Fe moments the Eu moments lie along the c-direction.

In this paper we present a systematic study of the interplay between magnetism and superconductivity in Eu$_{0.5}$K$_{0.5}$(Fe$_{1-x}$Ni$_{x}$)$_{2}$As$_{2}$. Since Ni-doping in antiferromagnetic EuFe$_{2}$As$_{2}$ is reported to suppress the SDW transition and results in a ferromagnetic order without any signature of superconductivity down to 2 K,\cite{21} it is of interest to see how Ni-doping modifies the superconducting ground state of Eu$_{0.5}$K$_{0.5}$Fe$_{2}$As$_{2}$. We therefore have investigated the effect of Ni-doping on the superconducting compound Eu$_{0.5}$K$_{0.5}$Fe$_{2}$As$_{2}$. We see that with an increase in Ni concentration the superconducting transition shifts rapidly to lower temperature and is completely suppressed (down to 1.8 K) for 8\% Ni-doping. Further we see that the spin density wave state reappears for $x$ = 0.08 and 0.12 samples as a result of the compensation of holes (K-doping at Eu-site) by the electrons (Ni-doping at Fe site). We discuss below our interesting observations of the coexistence and competition of magnetic order and superconductivity, and the emergence of antiferromagnetic ground state of Eu$^{2+}$ moments in Eu$_{0.5}$K$_{0.5}$(Fe$_{1-x}$Ni$_{x}$)$_{2}$As$_{2}$.

\section{EXPERIMENTAL DETAILS}

We have prepared the polycrystalline samples of Eu$_{0.5}$K$_{0.5}$(Fe$_{1-x}$Ni$_{x}$)$_{2}$As$_{2}$ ($x$ = 0, 0.03, 0.05, 0.08 and 0.12) using solid state reaction. The sample preparation process was similar to that described in Ref.~16. High purity elements taken in stoichiometric ratio were placed inside an alumina crucible which was sealed in a tantalum crucible. The tantalum crucible was then sealed inside an evacuated quartz ampoule and subsequently kept into the furnace for first heat treatment. The furnace was ramped slowly to 600$^\circ$C at a rate of 50$^\circ$C/h and kept there for 12 h to prereact As and  then heated to 900$^\circ$C where it was kept for 2 days. The prereacted products were ground thoroughly and pressed into pellets and annealed at 900$^\circ$C for 5 days. The sample handling was done inside a glove box having high purity argon atmosphere. The samples were characterized by powder x-ray diffraction with Cu $K_\alpha$ radiation to determine the phase purity and crystal structure. Scanning electron microscope (SEM) equipped with energy dispersive x-ray (EDX) analysis was used to check the homogeneity and composition of the sample. The magnetization measurements were performed using the superconducting quantum interference device (SQUID) magnetometer (MPMS, Quantum design). Resistivity and magnetoresistance measurements were carried out using standard four probe technique in a liquid helium cryogen and using the physical properties measurement system (PPMS, Quantum design). The heat capacity measurements were performed using the thermal relaxation method in the PPMS.

\section{RESULTS AND DISCUSSION}

\begin{figure}
\begin{center}
\includegraphics [width=8.5cm]{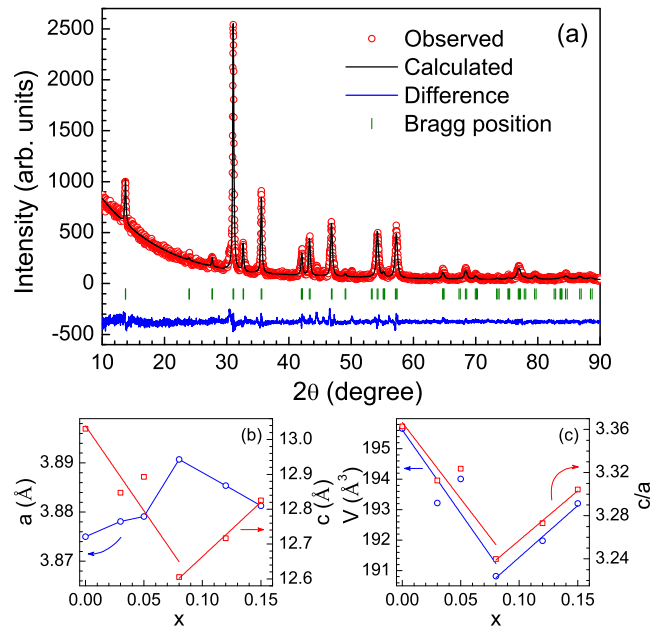}
\end{center}
\caption{\label{Fig 1} (Color online) (a) The powder X-ray diffraction pattern of Eu$_{0.5}$K$_{0.5}$(Fe$_{0.95}$Ni$_{0.05}$)$_{2}$As$_{2}$ recorded at room temperature. The solid line through the experimental points is the Rietveld refinement profile calculated for ThCr$_{2}$Si$_{2}$-type tetragonal (space group $I4/mmm$) structural model. The short vertical bars indicate the Bragg peak positions and the lowermost curve represents the difference between the experimental and model results. (b) The plot of lattice parameters $a$ and $c$ as a function of concentration $x$. (c) The variation of unit-cell volume $V$ and $c/a$ ratio as a function of $x$.}
\end{figure}

The room temperature x-ray diffraction (XRD) data obtained on the powdered samples of Eu$_{0.5}$K$_{0.5}$(Fe$_{1-x}$Ni$_{x}$)$_{2}$As$_{2}$ ($x$ = 0.03, 0.05, 0.08 and 0.12) indicate that all the compounds crystallize in ThCr$_{2}$Si$_{2}$-type tetragonal crystal structure (space group $I4/mmm$). The XRD data were analyzed by Rietveld refinement using Fullprof software which is shown in Fig.~1 for Eu$_{0.5}$K$_{0.5}$(Fe$_{0.95}$Ni$_{0.05}$)$_{2}$As$_{2}$. The lattice parameters, unit cell volumes and the position of arsenic atom $z_{As}$ obtained for all the samples are listed in Table~I. Fig.~1~(b) and (c) show the plots of lattice parameters $a$, $c$ and the unit-cell volume $V$, ratio $c/a$ as a function of Ni concentration $x$, respectively. It is seen from Fig.~\ref{Fig 1}(b) that with the increase in Ni concentration the lattice parameter $a$ initially increases up to $x$ = 0.08 beyond which it shows a decrease while the lattice parameter $c$ decreases up to $x$ = 0.08 (except a slight deviation for $x = 0.05$) and increases with further increase in $x$. The reason for the small deviation of lattice parameter $c$ for $x$ = 0.05 sample from the observed behavior is not clear. However, both unit-cell volume $V$ and $c/a$ ratio initially decrease with increasing $x$ up to $x = 0.08$, above which they start increasing as $x$ is increased further, at least up to $x=0.15$ (see Fig.~\ref{Fig 1}(c)). The impurity phase(s) estimated in our samples from the un-indexed XRD peaks and SEM images is less than 4\%. The composition and homogeneity of the samples were confirmed by EDX measurements.

\begin{table}
\caption{\label{Table I} Lattice parameters $a$ and $c$, $c/a$ ratio, unit-cell volume $V$ and the position of arsenic atom $z_{As}$ of ThCr$_{2}$Si$_{2}$-type tetragonal system Eu$_{0.5}$K$_{0.5}$(Fe$_{1-x}$Ni$_{x}$)$_{2}$As$_{2}$ ($x$ = 0, 0.03, 0.05, 0.08, 0.12 and 0.15).}
\begin{ruledtabular}
\begin{tabular}{c c c c c c}
$x$ & $a$ ({\AA})  &$c$ ({\AA})		&$c/a$		& $V$ ({\AA$^3$}) 	& $z_{As}$ \\[0.5ex]
\hline
0 	 & 3.8750(2)	&13.0309(2)		&3.36		& 195.66(1) 	& 0.3560(1) \\[1ex]

0.03 & 3.8781(1)	&12.8472(3)		&3.31		& 193.20(2) 	& 0.3562(1) \\[1ex]

0.05 & 3.8791(3)	&12.8931(1)		&3.32		& 193.99(1) 	& 0.3588(2)\\[1ex]

0.08 & 3.8907(2)	&12.6054(2)		&3.24		& 190.81(1) 	& 0.3619(1) \\[1ex]

0.12 & 3.8854(2)	&12.7170(1)		&3.28		& 191.96(2) 	& 0.3700(1) \\[1ex]

0.15 & 3.8813(3)	&12.8255(3)		&3.31		& 193.20(1) 	& 0.3761(2) \\
\end{tabular}
\end{ruledtabular}
\end{table}

\subsection{MAGNETIZATION}

\begin{figure}
\begin{center}
\includegraphics [width=8.5cm]{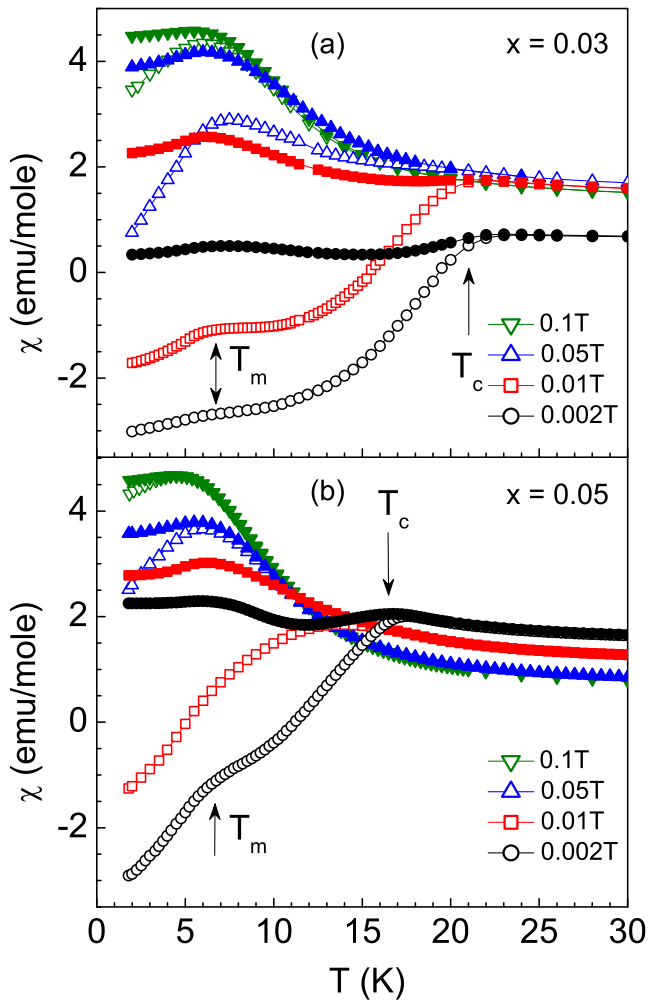}
\end{center}
\caption{\label{Fig 2} (Color online) (a) The temperature dependence of zero field cooled (ZFC) (open symbols) and field cooled (FC) (closed symbols) dc magnetic susceptibility of Eu$_{0.5}$K$_{0.5}$(Fe$_{0.97}$Ni$_{0.03}$)$_{2}$As$_{2}$ measured under different applied magnetic fields. (b) ZFC and FC data of Eu$_{0.5}$K$_{0.5}$(Fe$_{0.95}$Ni$_{0.05}$)$_{2}$As$_{2}$ sample  for various magnetic fields. The superconducting and magnetic transitions are marked by arrows.}
\end{figure}

\begin{figure}
\begin{center}
\includegraphics [width=8.5cm]{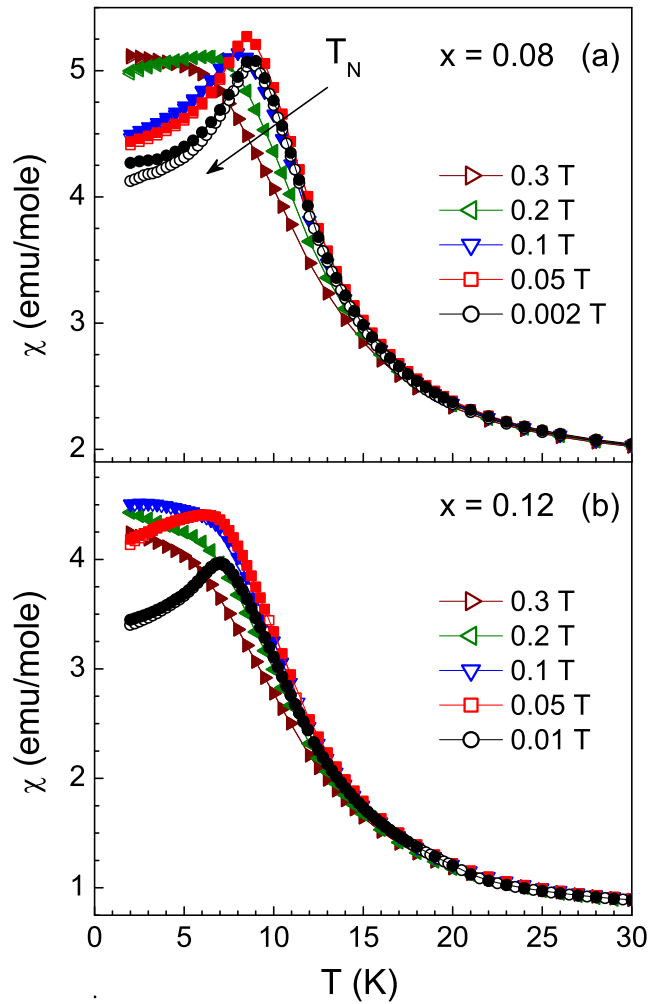}
\end{center}
\caption{\label{Fig 3} (Color online) (a) The temperature dependence of zero field cooled (ZFC) (open symbols) and field cooled (FC) (closed symbols) dc magnetic susceptibility of $x$ = 0.08 sample measured under different applied fields. (b) ZFC and FC data of $x$ = 0.12 sample for various magnetic fields. Note: the ZFC and FC data overlap, therefore the open symbols are not visible.}
\end{figure}

Figure 2 shows the dc magnetic susceptibility, $\chi (T)$ data for Eu$_{0.5}$K$_{0.5}$(Fe$_{1-x}$Ni$_{x}$)$_{2}$As$_{2}$ ($x$ = 0.03 and 0.05) measured under various applied magnetic fields. The parent compound Eu$_{0.5}$K$_{0.5}$Fe$_{2}$As$_{2}$ is reported to exhibit a superconducting transition below 33 K.\cite{16,23} The Ni-doping results in a rapid reduction in the superconducting transition temperature $T_{c}$. 3\% of Ni is found to suppress $T_{c}$ from 33 K to 23 K, and 8\% of Ni completely destroys the superconductivity (down to 1.8 K). Fig. 2(a) shows the magnetic susceptibility data of 3\% Ni-doped compound Eu$_{0.5}$K$_{0.5}$(Fe$_{0.9}$$_{7}$Ni$_{0.0}$$_{3}$)$_{2}$As$_{2}$. Zero field cooled (ZFC) dc magnetic susceptibility (open symbols) measured at 2~mT exhibits an onset of superconductivity at 23 K with a broad re-entrant type hump at 6 K, below which the diamagnetic signal increases again. This broad hump at 6 K which is due to Eu short range magnetic ordering is more pronounced in the ZFC $\chi (T)$ data measured under an applied field of 10~mT in comparison with the $\chi (T)$ data at 2~mT. The field cooled (FC) curve (closed symbols) also shows a decrease at 23 K due to magnetic flux expulsion in the superconducting state with a sudden increase below 10 K due to Eu$^{2+}$ magnetic ordering. While the onset of superconducting transition temperature decreases and finally suppressed below 1.8 K, the hump at 10 K becomes more pronounced and turns into a broad peak which further saturates for the fields above 0.1 T. 

The interplay of Eu magnetic ordering with superconductivity is also evident for $x$ = 0.05 as shown in the Fig. 2(b), where we see a  distinct broad shoulder from 10 to 6 K in ZFC $\chi (T)$ and the superconducting transition temperature is also decreased to 18 K. An increase in Ni concentration leads to further decrease in the superconducting transition temperature and superconductivity is not observed down to 1.8 K for $x$ = 0.08 (Fig. 3(a)). As is evident from the Fig. 3(a) there is no signature of superconductivity in ZFC $\chi (T)$ data of $x$ = 0.08 sample, however $\chi (T)$ is observed to exhibit a peak at 8.5 K; there is almost no hysteresis in ZFC and FC data which suggests that the transition is associated with the magnetic ordering of Eu$^{2+}$ moments and not due to any spin-glass freezing. The position of this peak depends on the magnetic field and a small shift to the lower temperature side (e.g. to 7.6 K at 0.2 T from 8.5 K at 0.002 T) is observed with an increase in the magnetic field which is a characteristic of antiferromagnetic ordering, thus suggesting an antiferromagnetic transition at $T_{N}$ = 8.5 K in 8\% Ni-doped sample. The magnetic susceptibility tends to become constant at an applied field of 0.2 T and above. The observation of shift of magnetic ordering temperature towards the lower temperature in Eu$_{0.5}$K$_{0.5}$(Fe$_{0.92}$Ni$_{0.}$$_{0}$$_{8}$)$_{2}$As$_{2}$ suggests the emergence of an antiferromagnetic ground state in contrast to the emergence of ferromagnetic ordering of Eu$^{2+}$ moments reported earlier with Ni-doping in EuFe$_{2}$As$_{2}$.\cite{21} The 12\% Ni-doped compound Eu$_{0.5}$K$_{0.5}$(Fe$_{0.88}$Ni$_{0.12}$)$_{2}$As$_{2}$ also orders antiferromagnetically below 7 K as shown in Fig. 3(b) and the features are similar to those observed for the sample with $x$ = 0.08. The high temperature magnetic susceptibility of all the samples follows the Curie-Weiss behavior (data not shown) with an effective magnetic moment close to 5.6 to 6$\mu_{B}$/f.u. where f.u. stands for formula unit.

\begin{figure}
\begin{center}
\includegraphics [width=8.5cm]{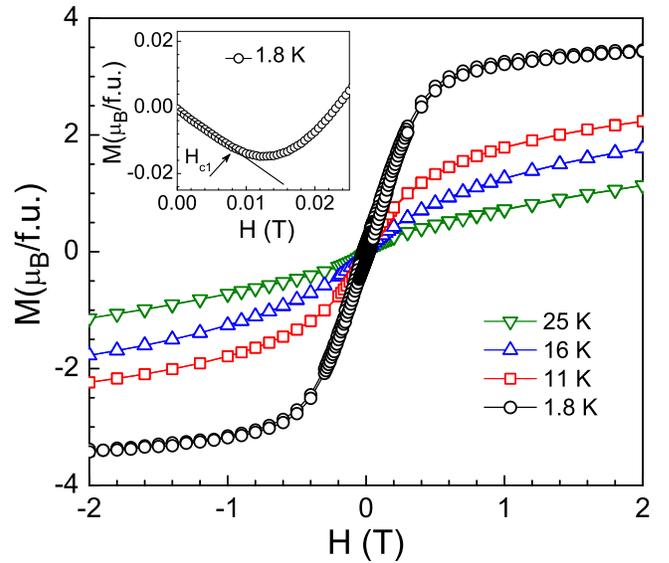}
\end{center}
\caption{\label{Fig 4} (Color online) The magnetic field dependence of the isothermal magnetization of Eu$_{0.5}$K$_{0.5}$(Fe$_{0.95}$Ni$_{0.05}$)$_{2}$As$_{2}$ at various temperatures. The inset shows the low field magnetization at 1.8 K}
\end{figure}

Figure 4 shows the field dependence of the isothermal magnetization for Eu$_{0.5}$K$_{0.5}$(Fe$_{0.95}$Ni$_{0.05}$)$_{2}$As$_{2}$ at various temperatures. The $M(H)$ curves below the superconducting transition temperature are composed of two components, one due to the superconductivity (diamagnetic) and other due to the paramagnetic component of Eu$^{2+}$ moments. The inset of Fig. 4 shows the $M(H)$ curve at 1.8 K at low fields. We clearly see the diamagnetic signal due to the Meissner effect at low fields. The deviation of the $M(H)$ curve from the Meissner line gives a rough estimate of the lower critical field, $H_{c1}$ which is $\approx$ 8~mT. At higher fields the $M(H)$ initially increases very rapidly and eventually tends to saturate above 1 T. The saturation magnetization is $\approx$ 7$\mu_{B}$/Eu indicating the divalent valence state of Eu in this compound. A very narrow hysteresis is also observed in $M(H)$ curve at 1.8 K. With an increase in temperature the lower critical field decreases and at 25 K which is well above the superconducting transition temperature, the magnetization varies almost linearly with field.

\begin{figure}
\begin{center}
\includegraphics [width=8.5cm]{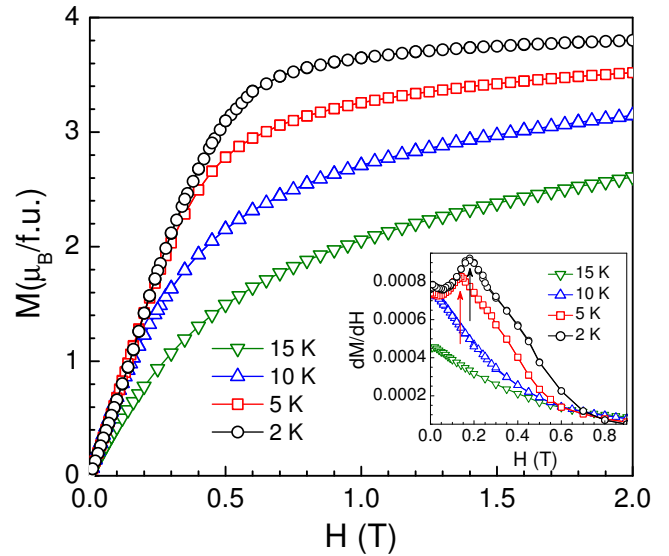}
\end{center}
\caption{\label{Fig 5} (Color online) The magnetic field dependence of the isothermal magnetization of Eu$_{0.5}$K$_{0.5}$(Fe$_{0.92}$Ni$_{0.08}$)$_{2}$As$_{2}$ at various temperatures. The inset shows the derivative of magnetization curves at 2, 5, 10 and 15 K.}
\end{figure}

The isothermal magnetization of $x$ = 0.08 sample shows only a very small change in slope at 0.18 T in the magnetically ordered state (e.g., at 2 K),  suggesting a weak  spin reorientation, and attains a value of $\sim$  3.5 $\mu_{B}$/f.u. at 0.7 T and tends to saturate with further increase in field (Fig.~5). At 2~K and 2~T the saturation moment is $\sim$ 3.7 $\mu_{B}$/f.u. ($\approx$ 7.4 $\mu_{B}$/Eu ion) which is slightly higher than the expected theoretical saturation magnetization $M_{s}$ = 7 $\mu_{B}$ for Eu$^{2+}$ ions ($J$ = 7/2). The small change in slope is apparent from the derivative plot which shows a peak at 0.18 T in 2 K curve (inset of Fig.~5). The position of this peak moves towards the lower field at 5~K and disappears completely at 10 and 15~K. In contrast to pure EuFe$_{2}$As$_{2}$, we do not observe any field induced step-like metamagnetic transition in our sample. It is important to mention that the helical magnetic ordering of Eu ions has been proposed to account for the destruction of superconductivity and hence the resistivity re-entrance in Eu(Fe$_{0.89}$Co$_{0.11}$)$_{2}$As$_{2}$.\cite{20} The helical magnetic structure in Eu(Fe$_{0.89}$Co$_{0.11}$)$_{2}$As$_{2}$ has been confirmed from the M\"{o}ssbauer spectroscopic studies.\cite{22} Magnetization measurements on Ni-doped EuFe$_{2}$As$_{2}$ inferred a ferromagnetic ordering of Eu ions below 18 K\cite{21}, however the M\"{o}ssbauer spectroscopic studies points towards a helical ordered state of Eu ions, the Eu spins are tilted at an angle of 36$^\circ$  from the crystallographic $c$-axis.\cite{22} Thus the absence of step like metamagnetic transition in our sample which is usually associated with the spin flip process in the A-type antiferromagnetism as seen in EuFe$_{2}$As$_{2}$, and a small shift of magnetic ordering temperature towards the lower temperature with magnetic field points towards a non-collinear arrangement of Eu$^{2+}$ spins and hence a helical magnetic ordering.

Thus we see that the magnetic ordering of Eu ions plays an important role in the suppression of superconductivity in Eu$_{0.5}$K$_{0.5}$(Fe$_{1-x}$Ni$_{x}$)$_{2}$As$_{2}$. Even though the Eu ions are diluted by 50\% potassium doping we see that the Eu magnetic ordering is very detrimental to the superconductivity. For low Ni concentration the Eu magnetic ordering coexists with superconductivity, however as the magnetic ordering temperature of Eu ions become greater than the superconducting $T_{c}$, it hinders the formation of Cooper pairing and destroys the superconductivity. Similar behavior has been reported for EuFe$_{1.9}$Co$_{0.1}$As$_{2}$ under the external hydrostatic pressure.\cite{24} In EuFe$_{1.9}$Co$_{0.1}$As$_{2}$ the Eu magnetic ordering temperature is enhanced with increasing pressure whereas the superconducting $T_{c}$ decreases and once the magnetic ordering temperature becomes greater than the superconducting $T_{c}$, the superconductivity gets suppressed below 1.8 K. The isothermal magnetization data obtained at 2 K for $x$ = 0.12 sample (data not shown) also exhibits a very small change in slope at 0.07 T pointing towards a weak spin reorientation, and attains a value of $\sim$  3.5 $\mu_{B}$/f.u. at 0.7 T and tends to saturate with further increase in field.

\begin{figure}
\begin{center}
\includegraphics [width=8.5cm]{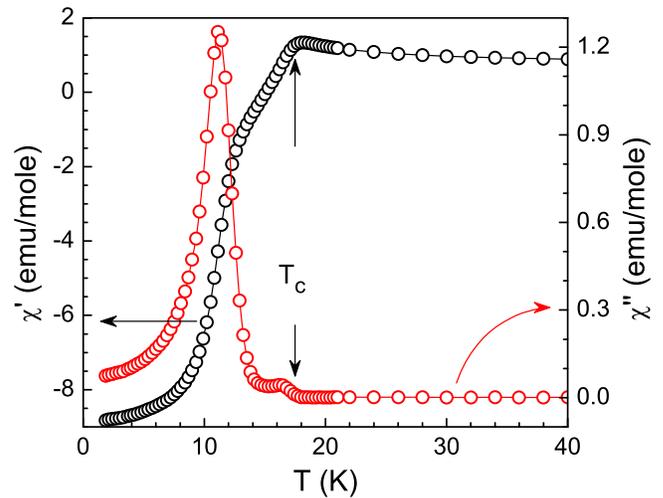}
\end{center}
\caption{\label{Fig 6} (Color online) The real and imaginary part of ac magnetic susceptibility for Eu$_{0.5}$K$_{0.5}$(Fe$_{0.95}$Ni$_{0.05}$)$_{2}$As$_{2}$ at a frequency of 12 Hz in an excitation field of 1 Oe.}
\end{figure}

The real and imaginary part of ac magnetic susceptibility of Eu$_{0.5}$K$_{0.5}$(Fe$_{0.95}$Ni$_{0.05}$)$_{2}$As$_{2}$ measured under an excitation field of 1 Oe and frequency of 12 Hz is shown in Fig. 6. The onset of superconducting transition at 18 K is clearly observed by a decrease in the real part of ac magnetic susceptibility ($\chi'$) due to shielding. Another anomaly is observed at 12.5 K in $\chi'$ below which we see a sharp increase in the diamagnetic signal. Thus the superconducting state seems to be weakened by Eu moment fluctuations just below $T_{c}$, which disappears below 10 K leading to a large increase of the diamagnetic signal. Correspondingly, the imaginary part of ac magnetic susceptibility ($\chi''$) exhibits a small hump near 17 K corresponding to the onset of superconducting transition. We also observe a sharp rise below 12 K with a peak at 10.5 K which could be due to the re-entrant superconductivity.

\subsection{RESISTIVITY AND MAGNETORESISTIVITY}

\begin{figure}
\begin{center}
\includegraphics [width=8cm]{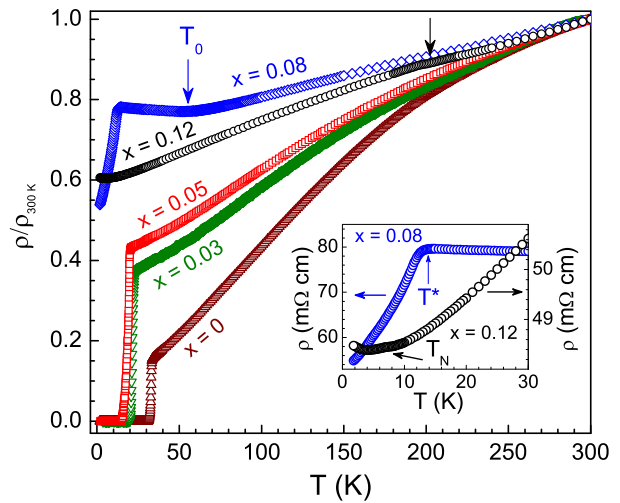}
\end{center}
\caption{\label{Fig 7} (Color online) The temperature dependence of the electrical resistivity of  Eu$_{0.5}$K$_{0.5}$(Fe$_{1-x}$Ni$_{x}$)$_{2}$As$_{2}$  ($x$ = 0, 0.03, 0.05, 0.08, 0.12) measured in  zero field. The inset shows the low temperature resistivity for $x$ = 0.08 and 0.12 sample below 30 K.}
\end{figure}

Figure 7 represents the temperature dependence of normalized electrical resistivity $\rho$(T)/$\rho_{300K}$ for Eu$_{0.5}$K$_{0.5}$(Fe$_{1-x}$Ni$_{x}$)$_{2}$As$_{2}$ ($x$ = 0, 0.03, 0.05, 0.08 and 0.12). The residual resistivities ($\rho_{0}$) for the superconducting samples just above the superconducting transition are $\rho_{0}$= 6.3, 0.68, and 0.42 m$\Omega$ cm for $x$ = 0, 0.03, and 0.05, respectively. However, the value of the residual resistivities for $x$ = 0.08 and 0.12 sample are $\rho_{0}$ $\approx$ 54 and 46 m$\Omega$ cm, respectively. While the high-temperature resistivity exhibits a metallic behavior, sharp superconducting transitions are observed at 33, 23 and 18 K in $x$ = 0, 0.03 and 0.05 samples, respectively. Thus, the resistivity data also confirm that an increase in Ni concentration results in a decrease in superconducting transition temperature ($T_{c}$) and eventually leads to complete suppression (at least down to 1.8 K) of superconductivity for $x$ = 0.08. The resistivity for $x$ = 0.08 exhibits a metallic behavior down to 100 K followed by an upturn at around 60 K and then a peak at 13 K below which resistivity decreases again. The most interesting and surprising feature in the resistivity is the upturn at $T_{0}$ ($\approx$ 60 K) and the peak at $T^*$ ($\approx$ 13 K) which is well above the antiferromagnetic ordering temperature ($T_{N}$ = 8.5 K) of Eu ions as obtained from the magnetic susceptibility and specific heat (described below) measurements. Similar upturn as seen at 60 K has been observed in the under doped compounds EuFe$_{2-x}$Ni$_{x}$As$_{2}$ ($x$ = 0.06, 0.09) and attributed to the spin density wave anomaly.\cite{21} We propose that this upturn observed in the resistivity data of Eu$_{0.5}$K$_{0.5}$(Fe$_{0.92}$Ni$_{0.08}$)$_{2}$As$_{2}$ could be due to the reappearance of the SDW in the system. Since the replacement of Eu by K in EuFe$_{2}$As$_{2}$ leads to hole doping while Ni-doping at Fe site leads to electron doping, therefore the electron doping in Eu$_{0.5}$K$_{0.5}$(Fe$_{1-x}$Ni$_{x}$)$_{2}$As$_{2}$ gradually compensate the optimal K-doping. This gives rise to the reappearance of spin density wave anomaly for $x$ = 0.08. We see that with further increase in Ni concentration, $x$ = 0.12, the anomaly due to the SDW shifts to a high temperature and occurs at $T_{0}$ = 200 K. For this concentration the electron doping fully compensates for the holes and the SDW transition occurs at the same temperature as observed in parent compound EuFe$_{2}$As$_{2}$.\cite{15} Thus we see that the physical properties of the system are strongly controlled by the charge on the FeAs layer. The tuning of superconducting to magnetic state and back to the superconducting state has also been reported for Ba$_{1-x}$K$_{x}$Fe$_{1.86}$Co$_{0.14}$As$_{2}$.\cite{25} Now we turn our discussion to the peak observed at $T^*$ $\approx$ 13 K for $x$ = 0.08. Similar transition has been observed in NdFeAsO where the antiferromagnetic to ferromagnetic arrangement of Fe moments along the $c$-axis occurs at 15 K accompanied by the onset of Nd AFM order below 6 K.\cite{26} The peak observed in our sample may have the similar origin. We do not see any feature of this transition in the magnetic measurements due to overwhelming contribution of Eu paramagnetism. However, the resistivity is quite sensitive to change in scattering rate and the observed transition is likely due to the changes in the magnetic scattering rate of the charge carriers. It seems that the Eu ion moments in this compound influence the Fe sublattice ordering in the adjacent FeAs layer and hence the antiferromagnetic ordering of Fe moments change to ferromagnetic along the $c$-axis. However, we do not observe any such features for $x$ = 0.12 as shown in the inset of Fig. 7. Instead we see a kink associated with the magnetic ordering of Eu ions at 7 K followed by an increase in the resistivity down to 1.8 K which could be due to an increase in disorder.

\begin{figure}
\begin{center}
\includegraphics [width=8cm]{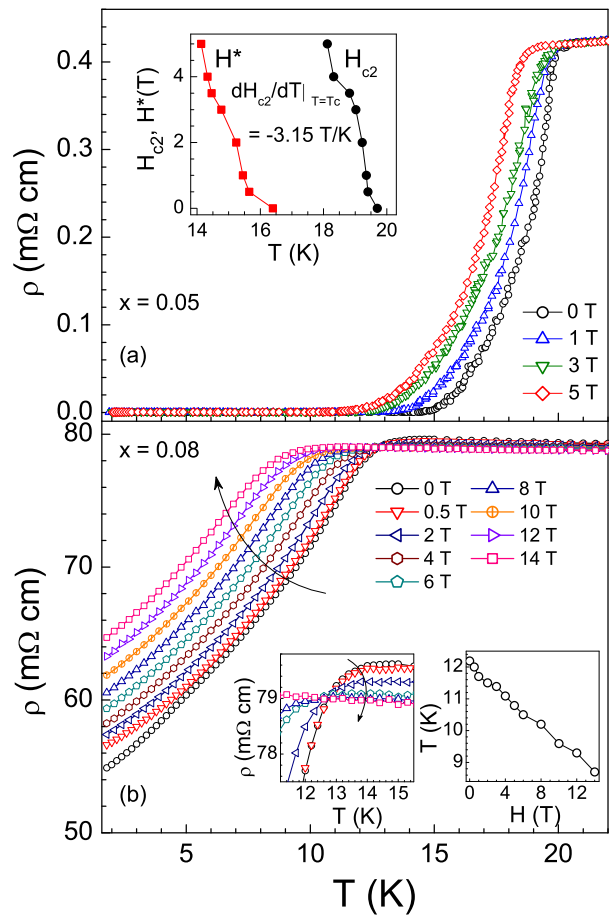}
\end{center}
\caption{\label{Fig 8} (Color online) (a) The temperature dependence of the electrical resistivity of Eu$_{0.5}$K$_{0.5}$(Fe$_{0.95}$Ni$_{0.05}$)$_{2}$As$_{2}$ at various applied magnetic fields. The inset shows the $H_{c2}$(T) and $H^{*}-T$ phase diagram. (b) The temperature dependence of electrical resistivity of Eu$_{0.5}$K$_{0.5}$(Fe$_{0.92}$Ni$_{0.08}$)$_{2}$As$_{2}$ measured under various applied fields. The left inset shows the expanded view in the temperature range 11--16 K. The right inset shows the $H-T$ phase diagram.}
\end{figure}

The field dependence of electrical resistivity for Eu$_{0.5}$K$_{0.5}$(Fe$_{0.95}$Ni$_{0.05}$)$_{2}$As$_{2}$ as a function of temperature is presented in Fig. 8(a). It is seen that like other iron pnictide superconductors the SC is very robust to magnetic field in this system. We find that with the increase in the magnetic field, the onset of the superconducting transition shifts slowly towards the lower temperature. The decrease in the temperature at which the sample reaches zero resistance is more rapid than the onset of superconducting transition. We have calculated the upper critical field $H_{c2}$(T) and the irreversibility field $H^*$(T) based on the resistivity measurements using the 90\% and 10\% values of the normal state resistivity $\rho_{n}$ at $T$ $\sim$  $T_{c}$, respectively, at different magnetic fields. The obtained values of $H_{c2}$(T) and $H^*$(T) are plotted in the inset of Fig. 8(a). The slope of the upper critical field $H_{c2}$(T) is approximately equal to -3.15 T/K which is smaller than the value reported for Eu$_{0.5}$K$_{0.5}$Fe$_{2}$As$_{2}$ ($\approx$ -4.45 T/K).\cite{23}  The value of zero temperature upper critical field $H_{c2}$(0) calculated using the Werthamer-Helfand-Hohenberg (WHH) formula, $H_{c2}^{Orb}$(0) = -0.693\,$T_{c}$($dH_{c2}/dT$)$\mid_{T=T_c}$, comes out to be 39.3 T in the dirty limit\cite{27} and $H_{c2}^{Orb}$(0) = -0.727\,$T_{c}$($dH_{c2}/dT$)$\mid_{T=T_c}$ = 41.2 T in the clean limit.\cite{28} These values of $H_{c2}$ are higher than the Pauli-Clogston limiting field, $H_{P}$ = 1.86 $T_{c}$ = 33.5 T.\cite{29} This suggests that the critical field $H_{c2}$ is essentially determined by the orbital pair breaking. Further, the value of Maki parameter $\alpha$ estimated using $\alpha_{M}$ = $\surd$2$H_{c2}$/$H_{P}$ = 1.66, which is high and might be an indication of the existence of Fermi pockets. The mean-field Ginzburg-Landau coherence length estimated using the relation $\xi$ = ($\phi_{0}$/2$\pi$$H_{c2}$)$^{1/2}$ ($\phi_{0}$ = 2.07 $\times$ 10$^{-11}$ T cm$^{2}$ being the flux quantum) is 27 {\AA} which is larger than that reported for Eu$_{0.5}$K$_{0.5}$Fe$_{2}$As$_{2}$ ($\approx$ 18 {\AA}).\cite{23}

To gain more insight into the transition observed at 13 K for Eu$_{0.5}$K$_{0.5}$(Fe$_{0.92}$Ni$_{0.08}$)$_{2}$As$_{2}$, we performed the field dependent resistivity measurements. Figure 8(b) shows the temperature dependence of the electrical resistivity of Eu$_{0.5}$K$_{0.5}$(Fe$_{0.92}$Ni$_{0.08}$)$_{2}$As$_{2}$ measured under various applied magnetic fields. We see that with the increase in the magnetic field the resistivity increases at low temperature which could be due to an increase in spin disorder scattering in the antiferromagnetically ordered state of Eu moments. As we move further towards the higher temperature, we see that at a particular field the resistivity coincides with the zero field resistivity value at around $T$ $\approx$ 13 K and decreases beyond with further increase in temperature. This means that the magnetoresistance practically comes down to zero at a particular temperature and field. This feature of resistivity data is more clear in the expanded view of the resistivity plot in the temperature range 12 to 15 K shown in the left inset of Fig. 8(b). The right inset of Fig. 8(b) presents the  $H-T$ phase diagram which clearly shows that the onset of the transition observed at $T^*$ = 13 K shifts towards the lower temperature with increasing  magnetic field.

\begin{figure}
\begin{center}
\includegraphics [width=8cm]{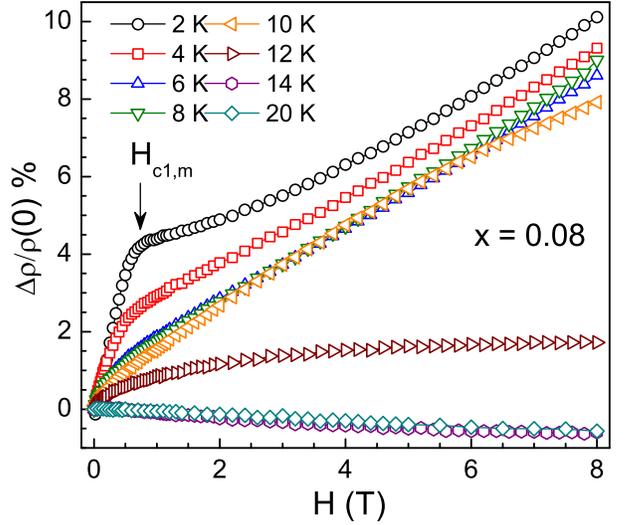}
\end{center}
\caption{\label{Fig 9} (Color online) The magnetic field dependence of the normalized magnetoresistance $\Delta$$\rho$/$\rho$(0) for Eu$_{0.5}$K$_{0.5}$(Fe$_{0.92}$Ni$_{0.08}$)$_{2}$As$_{2}$ at various temperatures.}
\end{figure}

\begin{figure}
\begin{center}
\includegraphics [width=8cm]{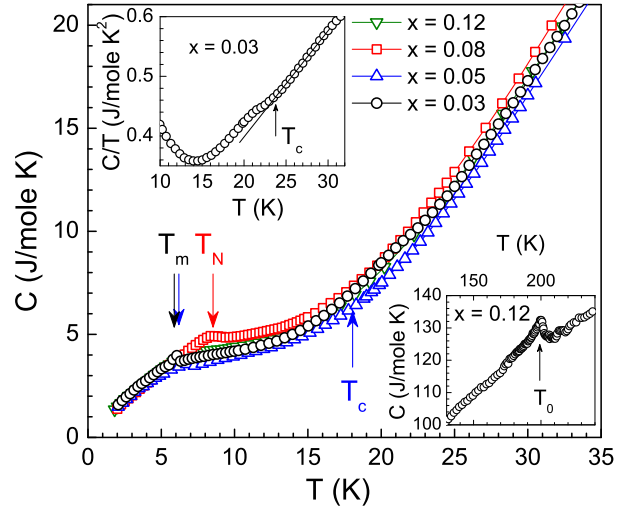}
\end{center}
\caption{\label{Fig 10} (Color online) The temperature dependence of specific heat of Eu$_{0.5}$K$_{0.5}$(Fe$_{1-x}$Ni$_{x}$)$_{2}$As$_{2}$ ($x$ = 0.03, 0.05, 0.08 and 0.12) measured under zero magnetic field. The superconducting and magnetic transitions are marked by arrows. The upper inset shows the C/T vs. T plot for $x$ = 0.03 sample. The superconducting transition is  marked by arrow. The lower inset shows the specific heat for $x$ = 0.12 in temperature range 130--240 K.}
\end{figure}

Figure 9 depicts the field dependence of the resistivity plotted as normalized magnetoresistance $\Delta$$\rho$/$\rho_{0}$ [=($\rho$(H)-$\rho$(0))/$\rho$(0)] for Eu$_{0.5}$K$_{0.5}$(Fe$_{0.92}$Ni$_{0.08}$)$_{2}$As$_{2}$ at various temperatures. We see a moderate magnetoresistance of $\approx$ 10\% at 2 K under an applied field of 8 T. At 2 K which is well below the magnetic ordering temperature of Eu moments, the magnetoresistance increases linearly with increase in magnetic field due to the increase in spin disorder scattering as mentioned earlier. The kink at the critical magnetic field $H_{c1_,m}$ ($\approx$ 0.7 T) corresponds to the magnetic saturation in $M(H)$ curve above which the Eu$^{2+}$ spins align ferromagnetically. However, instead of decrease in resistivity as one would have expected in ferromagnetically ordered state, we observe that the magnetoresistance keeps on increasing even with further increase in magnetic field. Similar increase in resistance has also been observed in the field induced ferromagnetic state of EuFe$_{2}$As$_{2}$.\cite{30} With the increase in temperature, the step at $H_{c1_,m}$ decreases and disappears completely at 10 K. Further increase in temperature leads to a decrease in the value of MR and at 14 and 20 K we observe a negative magnetoresistance. This unusual behavior of MR also demonstrates the interplay of Eu moments AFM with the Fe moments in the FeAs layer. Thus the competition between the AFM and FM phases of Eu and Fe spins, respectively, leads to zero magnetoresistance at 13 K. At lower temperature the AFM phase is dominating and hence the overall MR is positive. To understand the origin of transition observed at 13 K, further investigations by neutron scattering and NMR measurements are highly desired.

\subsection{SPECIFIC HEAT}

\begin{table}
\caption{\label{Table II} The Sommerfeld coefficient $\gamma$ and the coefficient $\beta$ along with the Debye temperature $\Theta_{D}$ for the system Eu$_{0.5}$K$_{0.5}$(Fe$_{1-x}$Ni$_{x}$)$_{2}$As$_{2}$ ($x$ = 0, 0.03, 0.05, 0.08 and 0.12).}
\begin{ruledtabular}
\setlength{\tabcolsep}{8pt}
\begin{tabular}{c c c c}
$x$ & $\gamma$  			&$\beta$ 			&$\Theta_D$(K)\\
  &(mJ/mole K$^2$) 		&($\mu$J/mole K$^4$) \\[0.5ex]
\hline
0 	 & 260		&460			&276\\[1ex]

0.03 & 230		&496			&269\\[1ex]

0.05 & 187	 	&515			&265\\[1ex]

0.08 & 236		&532			&263 \\[1ex]

0.12 & 206		&562			&258\\
\end{tabular}
\end{ruledtabular}
\end{table}

The bulk nature of the superconducting and magnetic transitions is further confirmed by the specific heat measurements done in the temperature range 2 to 40 K. Figure 10 shows the specific heat of Eu$_{0.5}$K$_{0.5}$(Fe$_{1-x}$Ni$_{x}$)$_{2}$As$_{2}$ ($x$ = 0.03, 0.05, 0.08 and 0.12) measured under zero magnetic field. We see two anomalies in the specific heat data of $x$ = 0.03 sample at 23 and 6 K associated with the SC and Eu short range magnetic ordering, respectively, which are consistent with the aforementioned anomalies in the magnetic susceptibility data and hence indicates the coexistence of SC and short range magnetic ordering in this sample. The SC transition is more clear for $x$ = 0.03 from the $C/T$ vs. $T$ plot as shown in the upper inset of Fig. 10. However, the superconducting transition is not clearly visible for $x$ = 0.05 due to the influence of Eu magnetic ordering, while the anomaly corresponding to the magnetic ordering is clearly visible at $\approx$ 7 K. The specific heat of $x$ = 0.08 and 0.12 compounds also shows clear anomalies at 8.5 and 7 K, respectively, which are associated with the Eu magnetic ordering temperature $T_{N}$. The entropy associated with the transition is distributed even above the transition temperature in the wide temperature range up to 16 K as is clear from the specific heat data of $x$ = 0.08. The lower inset of Fig. 10 shows the specific heat data for $x$ = 0.12 in the temperature range 130 - 240 K. We clearly observe the anomaly associated with the spin density wave transition at 200 K as marked by the arrow and hence evidence the intrinsic nature of the transition. Further experiments like low temperature X-ray and M\"{o}ssbauer spectroscopic studies are required to see whether the spin density wave anomaly is accompanied and/or followed by the structural transition as observed for pure EuFe$_{2}$As$_{2}$.\cite{15} The specific heat data of all the compounds were fitted to the relation $C = \gamma T + \beta T^{3} + \delta T^{5}$ in the normal state i.e. in the temperature range 25-30 K for $x$ = 0.03 and 20-28 K for $x$ = 0.05, 0.08, 0.12 and the corresponding fitting parameters $\gamma$ and $\beta$ are enlisted in Table~II. The enhanced values of the Sommerfeld coefficient could be due to the influence of Eu magnetic ordering even above the transition temperature. We estimated the Debye temperature from the values of $\beta$ using the relation $\Theta_{D} = (12\pi^{4}N_{A}rk_{B}/5\beta)^{1/3}$, where $r$ is the number of atoms per formula unit. The obtained values of $\Theta_{D}$ are also shown in Table~II. We find that with the increase in the Ni concentration $x$, the Debye temperature decreases.

\subsection{FURTHER DISCUSSION}

We have seen that the increasing Ni concentration in Eu$_{0.5}$K$_{0.5}$(Fe$_{1-x}$Ni$_{x}$)$_{2}$As$_{2}$ leads to the destruction of superconductivity and an emergence of a long range magnetic ordering in Eu lattice. Apparently we are introducing site disorder in the FeAs superconducting layer by Ni substitution at Fe site and thus a peculiar question arises that why more disorder leads to long range magnetic ordering in Eu$_{0.5}$K$_{0.5}$(Fe$_{1-x}$Ni$_{x}$)$_{2}$As$_{2}$. One possibility is that a strong pair breaking effect occurs due to the disorder induced by the Ni substitution which leads to the SC decoupling of the layers. On the other hand there is a strong contraction of lattice along the $c$-axis as the $c/a$ ratio decreases by a factor of 3.6\% for $x$ = 0.08 as compared to $x$ = 0. Therefore the pair breaking due to the disorder induced by the Ni substitution in the FeAs layer and the lattice contraction along the c-axis modify the RKKY interaction and give rise to the long range magnetic ordering accompanied by complete destruction of superconductivity for $x$ = 0.08. The transition from the antiferromagnetic to paramagnetic phase via a superconducting phase has also been observed for a fully compensated state in Ba$_{1-x}$K$_{x}$(Fe$_{1-y}$Co$_{y}$)$_{2}$As$_{2}$ ($x$ = $2y$) and is attributed to the combined effect of the lattice contraction and the site disorder due to the Co$^{2+}$ substitution.\cite{31}

\section{CONCLUSION}

We have investigated the effect of Ni substitution on the magnetism and superconductivity in Eu$_{0.5}$K$_{0.5}$Fe$_{2}$As$_{2}$. The magnetic and transport properties examined by the various experimental techniques provide evidence for the coexistence and competition of magnetic order and superconductivity, and the emergence of antiferromagnetic ground state of Eu$^{2+}$ moments in Ni-doped Eu$_{0.5}$K$_{0.5}$Fe$_{2}$As$_{2}$ compounds. The superconducting transition temperature decreases with the increase in Ni concentration and SC is completely suppressed for 8\% Ni substitution. The Eu magnetic moment ordering strongly affects the superconductivity and $T_{c}$ is completely suppressed (at least down to 1.8 K) once the Eu magnetic ordering temperature $T_{m}$ becomes greater than $T_{c}$. The suppression of superconductivity with the increase in Ni concentration clearly demonstrates the important role of Eu magnetic ordering and its interplay with superconductivity. The small shift of Eu magnetic ordering temperature with magnetic field and the absence of field induced metamagnetic transition in the $M(H)$ curves for $x$ = 0.08 and 0.12 samples point towards the helical magnetic structure of the Eu magnetic moments. Thus the helimagnetism could be responsible for the destruction of the SC as has been proposed for the Co-doped EuFe$_{2}$As$_{2}$. The electrical resistivity and specific heat measurements clearly demonstrate the reappearance of spin density wave anomaly for $x$ = 0.08 and 0.12 which arises due to the compensation of the holes created on the K-doping by the electrons on Ni substitution for Fe.


\begin{thebibliography}{31}

\bibitem {1}

D. C. Johnston, Advances in Physics \textbf{59}, 803 (2010).

\bibitem {2}
Y. Maeno, T. Tomita, M. Kyogoku, S. Awaji, Y. Aoki, K. Hoshino, A. Minami, and T. Fujita, Nature \textbf{328}, 512 (1987).

\bibitem {3}
Y. O. Nakamura, N. Matsuda, and Y. Shiina, Sol. State Commun. \textbf{81}, 923 (1992).

\bibitem {4}
G. Xiao , M. Z. Cieplak, J. Q. Xiao, and C. L. Chien, Phys. Rev. B \textbf{42}, 8752 (1990).

\bibitem {5}
A. S. Sefat, M. A. McGuire, R. Jin, B. C. Sales, D. Mandrus, F. Ronning, E. D. Bauer, and Y. Mozharivskyj, Phys. Rev. B \textbf{79}, 094508 (2009).

\bibitem {6}
A. S. Sefat, R. Y. Jin, M. A. McGuire, B. C. Sales, D. J. Singh, and D. Mandrus, Phys. Rev. Lett. \textbf{101}, 117004 (2008).

\bibitem {7}
N. Ni, A. Thaler, J. Q. Yan, A. Kracher, E. Colombier, S. L. Bud'ko, P. C. Canfield, and S. T. Hannahs,  Phys. Rev. B \textbf{82}, 024519 (2010).

\bibitem {8}
A. Kreyssig, M. G. Kim, S. Nandi, D. K. Pratt, W. Tian, J. L. Zarestky, N. Ni, A. Thaler, S. L. Bud'ko, P. C. Canfield, R. J. McQueeney, and A. I. Goldman, Phys. Rev. B \textbf{81}, 134512 (2010).

\bibitem {9}
P. C. Canfield and S. L. Bud'ko, Annu. Rev. Condens. Matter Phys. \textbf{1}, 27 (2010).

\bibitem {10}
F. Han, X. Zhu, P. Cheng, G. Mu, Y. Jia, L. Fang, Y. L. Wang, H. Luo, B. Zeng, B. Shen, L. Shan, C. Ren, and H. H. Wen, Phys. Rev. B \textbf{80}, 024506 (2009).

\bibitem {11}
A. Leithe-Jasper, W. Schnelle, C. Geibel, and H. Rosner, Phys. Rev. Lett. \textbf{101}, 207004 (2008).

\bibitem {12}
S. R. Saha, N. P. Butch, K. Kirshenbaum, and J. Paglione, Phys. Rev. B \textbf{79}, 224519 (2009).

\bibitem {13}
Y. P. Qi, L. Wang, Z. S. Gao, D. L. Wang, X. P. Zhang, and Y. W. Ma, Physica C \textbf{469}, 1921 (2009).

\bibitem {14}
S. L. Bud'ko, N. Ni, and P. C. Canfield, Phys. Rev. B \textbf{79}, 220516R (2009).

\bibitem {15}
Z. Ren, Z. Zhu, S. Jiang, X. Xu, Q. Tao, C. Wang, C. Feng, G. Cao, and Z. Xu, Phys. Rev. B \textbf{78}, 052501 (2008).

\bibitem {16}
Anupam, P. L. Paulose, S. Ramakrishnan, and Z. Hossain, J. Phys.: Condens. Matter \textbf{23}, 455702 (2011).

\bibitem {17}
C. F. Miclea, M. Nicklas, H. S. Jeevan, D. Kasinathan, Z. Hossain, H. Rosner, P. Gegenwart, C. Geibel, and F. Steglich, Phys. Rev. B \textbf{79}, 212509 (2009).

\bibitem {18}
Z. Ren, Q. Tao, S. Jiang, C. Feng, C. Wang, J. Dai, G. Cao, and Z. Xu, Phys. Rev. Lett. \textbf{102}, 137002 (2009).

\bibitem {19}
H. S. Jeevan, D. Kasinathan, H. Rosner, and P. Gegenwart, Phys. Rev. B \textbf{83}, 054511 (2011).

\bibitem {20}
S. Jiang, H. Xing, G. Xuan, Z. Ren, C. Wang, Z. A. Xu, and G. Cao, Phys. Rev. B \textbf{80}, 184514 (2009).

\bibitem {21}
Z. Ren, X. Lin, Q. Tao, S. Jiang, Z. Zhu, C. Wang, G. Cao, and Z. Xu, Phys. Rev. B \textbf{79}, 094426 (2009).

\bibitem {22}
I. Nowik, I. Felner, Z. Ren, G. H. Cao, and Z. A. Xu, New J. Phys. \textbf{13}, 023033 (2011).

\bibitem {23}
Anupam, P. L. Paulose, H. S. Jeevan, C. Geibel, and Z. Hossain, J. Phys.: Condens. Matter \textbf{21}, 265701 (2009).

\bibitem {24}
M. Nicklas, M. Kumar, E. Lengyel, W. Schnelle, and A. Leithe-Jasper, J. Phys.: Conf. Ser. \textbf{273}, 012101 (2011).

\bibitem {25}
V. Zinth, T. Dellmann, H. H. Klauss, and D. Johrendt, Angew. Chem. Int. Ed. \textbf{50}, 1 (2011).

\bibitem {26}
W. Tian, W. Ratcliff, M. G. Kim, J. Q. Yan, P. A. Kienzle, Q. Huang, B. Jensen, K. W. Dennis, R. W. McCallum, T. A. Lograsso, R. J. McQueeney, A. I. Goldman, J. W. Lynn, and A. Kreyssig, Phys. Rev. B \textbf{82}, 060514(R) (2010).

\bibitem {27}
N. R. Werthame, E. Helfand, and P. C. Hohenber, Phys. Rev. \textbf{147}, 295 (1966).

\bibitem {28}
E. Helfand and N. R. Werthame, Phys. Rev. \textbf{147}, 288 (1966).

\bibitem {29}
A. M. Clogston, Phys. Rev. Lett. \textbf{9}, 266 (1962).

\bibitem {30}
S. Jiang, Y. K. Luo, Z. Ren, Z. W. Zhu, C. Wang, X. F. Xu, Q. Tao, G. H. Cao, and Z. A. Xu, New J. Phys. \textbf{11}, 025007 (2009).

\bibitem {31}
S. Suzuki, K. Ohgushi, Y. Kiuchi, and Y. Ueda, Phys. Rev. B \textbf{82}, 184510 (2010).


\end{thebibliography}
\end{document}